\RequirePackage{lineno} 

\documentclass[12pt,onecolumn,prc,showpacs,preprintnumber,amsmath,
floatfix,amssymb]{revtex4}
\usepackage{epsfig}
\usepackage{graphicx}
\usepackage{dcolumn}
\usepackage{bm}
\def    \nn             {\nonumber}
\def\var{\mbox{\boldmath $\varepsilon$}}
\def\r{\mbox{{\bf  r}}}
\def\p{\mbox{\boldmath $p$}}
\def\q{\mbox{\boldmath $q$}}
\def\k{\mbox{\boldmath $k$}}
\def\t{\mbox{\boldmath $t$}}

\allowdisplaybreaks[4]

\begin{document}

\title{
  Flux-integrated semiexclusive cross sections for charged-current
  quasielastic and neutral-current elastic neutrino scattering off ${}^{40}$Ar
  and a sterile neutrino oscillation study}
\author{A.~V.~Butkevich}
\affiliation{ 
  Institute for Nuclear Research, Russian Academy of Sciences, 60th October
  Anniversary Prosp. 7A, Moscow 117312, Russia \\
}
\date{\today}

\begin{abstract}

  Flux-integrated semiexclusive differential cross sections for
  charged-current quasielastic and neutral-current elastic neutrino scattering
  on argon are analyzed. The cross sections are calculated using the
  relativistic distorted-wave impulse approximation with values of the nucleon
  axial mass $M_A=1$ GeV and 1.2 GeV. The elastic scattering cross sections
  are also computed for different strange quark contributions to the
  neutral-current axial form factor. The flux-integrated differential cross
  sections as functions of reconstructed neutrino energy are evaluated for the
  far detector of the SBN experiment. The effects of the short base-line
  neutrino oscillations are taken into account in a 3+1 framework.
  We found that cross sections depend on oscillation
  parameters and the ratio of the measured and predicted cross sections can be
  used in a sterile neutrino oscillation study.
  
\end{abstract}

\maketitle

\section{Introduction}

Neutrino neutral current elastic (NCE) scattering off nuclei can give
information about the structure of the  hadronic weak neutral current (NC) and
on the strange quark contribution to the nucleon spin. In contrast, purely
isovector charged-current (CC) processes do not depend on the strange
form factors. Therefore the CC and NC processes give complementary information
on nuclear effects in neutrino-nucleus scattering.

The weak neutral current of the nucleon may be parameterized in terms of two 
vector and one axial-vector form factors. An additional induced pseudoscalar 
form factor is presented. The axial-vector form factor may be split into a 
non-strange and strange contributions. The latter is proportional to the
fraction of the nucleon spin carried by the strange
quarks~\cite{Alberico1,Garvey1}. The strange vector form factors were measured
in parity-violating electron scattering experiments~\cite{HAPPEX, SAMPLE, A41,
  A42, G0}. The combined analysis of these experimental data points to
small strangeness of the vector form factors ~\cite{Liu, Donnelly}. 
  
Neutrino induced reactions are sensitive to the strange quark contribution to
the NC axial-vector form factor. The strange axial form factor is parameterized
as a dipole and uses the same axial mass as applied for the non-strange
form factor; the strange axial coupling constant at four-momentum transfers
squared $Q^2=0$ is $\Delta s$.
A measurement of $\nu$($\bar{\nu}$)-proton NCE at Brookheven National
Laboratory (BNL E734)~\cite{BNL} suggested a non-zero value of $\Delta s$. 
However this result suffers strongly from experimental uncertainties due to
difficulties in determination of the absolute neutrino flux. The measurement
of the neutral-to-charged-current (CC) quasielastic (QE) cross section
$R=NCE/CCQE$ 
in neutrino-nucleus scattering was proposed in Ref.~\cite{FINeSSE} to 
extract information on the strange spin of the proton because much of the
systematic uncertainty is canceled by using the ratio. The MiniBooNE
experiment measured the flux-integrated NCE differential cross section
$d\sigma/dQ^2$ as a function  of four-momentum transferred squared $Q^2$ and
the ratio $R=NCE/CCQE$ to extract the value $\Delta s$~\cite{MiniB1}.

Over the last few decades, a series of anomalous neutrino flavor oscillation
measurements~\cite{LSND,MiniB2,Mention,Gallex,BEST} have been made at short
base-line that could be explained by the existence of one more eV-scale
neutrinos, the so-called ``sterile neutrino'' $\nu_s$. The sterile neutrino is
addition to 3-flavor model with the three active neutrino
$\nu_{active}=\{\nu_e,\nu_{\mu},\nu_{\tau}\}$ and has no coupling to
either the W$^{\pm}$ or Z$^{0}$ bosons. The simplest extension to this 3-flavor
model is referred to as the 3+1 model and introduces in addition to massive
neutrino $\nu_1, \nu_2, \nu_3$ a single new mass state, $\nu_4$, with a
corresponding sterile flavor state $\nu_s$. On the other hand neither
long-baseline oscillation experiments~\cite{Minos, IceCube, NOVA}, or the SNO
experiment with solar neutrinos~\cite{SNO} have found evidence for this sterile
neutrino.

Motivated by the need for a resolution to the short base-line anomalies, the
Short-Base line Neutrino Program (SBN) using the Booster Neutrino Beam
(BNB) at Fermilab was proposed~\cite{SBN}. This experiment consists of three
liquid argon time projection chambers (LArTPCs) located at the distances of
hundreds of meters from BNB targets: a near detector SBND; an intermediate
detector MicroBooNE; and a far detector ICARUS. The positions of these detectors
optimized for study neutrino oscillation with mass-squared deference
$\Delta m^2_{i4} \sim 1$ eV$^2$, where $i=1,2,3$. Recently the MicroBooNE
collaboration presented the result of a measurement of BNB $\nu_e$ interactions,
to study the excess of low energy interactions observed by the MiniBooNE
~\cite{MiniB2} collaboration. The results are found to be consistent with the
nominal $\nu_e$ rate expectations from BNB and no excess of $\nu_e$ events is
 observed~\cite{MicroB1, MicroB2}.

 The NC interaction can play an important role in oscillation experiments.
 Because the three active neutrinos couple to $Z^0$, the rate of NC events
 should be unaffected by the 3-flavor neutrino oscillations. Therefore in a
 sterile search based on the NC interaction, the signal is the disappearance of
 any active neutrinos creating a deficit in rate of NC events.

 To evaluate the oscillation parameters, the probabilities of neutrino
 oscillations as functions of neutrino energy are measured. The accuracy to
 which neutrino oscillation parameters can be extracted depends on the
 ability of experiments to determine the individual energy of detected neutrino.
 In the few-GeV neutrino energy regime corresponding to the BNB, the
 quasielastic scattering is the dominant interaction mode and the kinematics of
 the outgoing lepton or proton are sufficient for determing the neutrino
 energy.

 The semiexclusive reaction $\nu + A \rightarrow \nu + p + B$ is a good signal
 sample of neutrino NCE scattering off nuclei. The measurements of the
 proton energy and its angle with respect to direction of the incident neutrino
 determine the neutrino energy. The LArTPCs provide low tracking thresholds and
 precise energy and angular resolution for charged particles, improving
 neutrino energy estimation in NC interactions.

 In recent years many theoretical studies have been presented to improve our
 knowledge on the NCE neutrino-nucleus scattering~\cite{Alberico2, Barbado,
   Martinez, Nieves, Leitner, BAV1, Gonza1, Meucci, Gonza2, Rocco, Giusti}. 
 In the semiexclusive process $(\nu_{\mu}, \nu_{\mu} p)$m the neutrino removes a
 single intact nucleon from the nucleus without producing any additional
 particles. Understanding the interaction of neutrinos with argon nuclei is of
 particular importance, since neutrino oscillation experiments such as
 DUNE ~\cite{DUNE}
 and SBN~\cite{SBN} employ neutrino detectors using LArTPCs. Unfortunately the
 cross section data for the semiexclusive lepton scattering on argon in the
 relevant energy range are rather scare. There are only experimental data for
 2.2 GeV electron scattering on argon~\cite{JLab1, JLab2} and
 flux-integrated differential CCQE-like~\cite{MicroB3} and NCE~\cite{MicroB4}
 cross sections for $\nu_{\mu}{}^{40}$Ar scattering measured with the
 MicroBooNE detector. The total uncertainty of the MicroBooNe measured NCE
 cross sections ranging from 50\% to 100\% at high energies.

 In the first part of this work the flux-integrated differential cross sections
 of ${}^{40}$Ar$(\nu_{\mu}, \nu_{\mu} p)$ interaction are calculated with the
 relativistic distorted-wave impulse approximation (RDWIA)~\cite{Pick1, Pick2,
 Kelly1, Kelly2, Udias1, Meucci2}, using the
 BNB. The RDWIA takes into account the nuclear shell structure and final state
 interaction (FSI) of the ejected nucleon with the residual nucleus. In our
 approach~\cite{BAV2} the effects of the short-range nucleon-nucleon (NN)
 correlations leading to the appearance of a high-momentum and high-energy
 distribution in the target are estimated. This approach was successfully
 applied in Refs.~\cite{BAV2, BAV3, BAV4, BAV5, BAV6} for calculation of the
 quasielastic semiexclusive and inclusive cross sections for the electron and
 neutrino scattering on ${}^{12}$C, ${}^{16}$O, ${}^{40}$Ca, and ${}^{40}$Ar
 nuclei.

 In the second part of this article we calculate the flux-integrated
 differential cross sections for semiexclusive CCQE and NCE scattering as
 functions of reconstructed with kinematic method neutrino energy. We explore
 possible application of these cross sections, calculated for the near SBND
 and far ICARUS detectors, for the sterile neutrino search at SBN. If these
 cross sections are extracted with good accuracy, this method can be applied
 to probe sterile neutrino oscillation parameters. 

 The outline of this paper is the following. In Sec. II we present briefly the
 formalism for the NCE semiexclusive scattering process and basic aspects of
 the RDWIA approach. The flux-integrated double and single differential cross
 sections are presented and discussed in Sec. III. In Sec. IV we show how the
 flux-integrated cross sections can be used to search for sterile neutrino at
 SBN in context of the 3+1 model. Our conclusions are summarized in Sec. V.

 
\section{The formalism and model for the neutral-current elastic scattering}

In this section we consider the formalism for description of NCE exclusive
\begin{equation}\label{Eq.1}
\nu(k_i) + A(p_A)  \rightarrow \nu(k_f) + N(p_x) + B(p_B),      
\end{equation}

scattering off nuclei in the one-$Z^0$-boson exchange approximation. Here 
$k_i=(\varepsilon_i,\k_i)$ 
and $k_f=(\varepsilon_f,\k_f)$ are the initial and final lepton 
momenta, $p_A=(\varepsilon_A,\p_A)$, and $p_B=(\varepsilon_B,\p_B)$ are 
the initial and final target momenta, $p_x=(\varepsilon_x,\p_x)$ is the 
ejectile nucleon momentum, $q=(\omega,\q)$ is the momentum transfer carried by 
the virtual $Z^0$-boson, and $Q^2=-q^2=\q^2-\omega^2$ is the $Z^0$-boson 
virtuality. As the basic outline follows closely the CC formalism developed in 
Ref.~\cite{BAV2}, we present a brief review that focuses on those 
modifications that arise from the weak neutral current.
 
\subsection{NCE neutrino-nucleus semiexclusive cross section} 

In the laboratory frame, the differential cross section for the exclusive
(anti-)neutrino NCE scattering, in which only a single 
discrete state or narrow resonance of the target is excited, can be written as
\begin{equation}
\label{Eq.2}
\frac{d^5\sigma^{(nc)}}{d\varepsilon_f d\Omega_f d\Omega_x} = R
\frac{\vert\p_x\vert{\varepsilon}_x}{(2\pi)^5}\frac{\vert\k_f\vert}   
{\varepsilon_i} \frac{G^2}{2} L_{\mu \nu}^{(nc)}W^{\mu \nu (nc)},
\end{equation}
 where $\Omega_f$ is the solid angle for the lepton momentum, $\Omega_x$ is the
 solid angle for the ejectile nucleon momentum,
 $G \simeq 1.16639 \times 10^{-11}$~MeV$^{-2}$ is the Fermi constant,
 $L^{(nc)}_{\mu \nu}$ and $W^{(nc)}_{\mu \nu}$ are NC lepton and nuclear tensors,
 respectively, and $R$ is a recoil factor
  \begin{equation}\label{Rec}
R =\int d\varepsilon_x \delta(\varepsilon_x + \varepsilon_B - \omega -m_A)=
{\bigg\vert 1- \frac{\varepsilon_x}{\varepsilon_B}
\frac{\p_x\cdot \p_B}{\p_x\cdot \p_x}\bigg\vert}^{-1}. \nn                 
\end{equation}
The energy $\var_x$ is the solution to the equation
\begin{equation}\label{Eq.4}
\varepsilon_x+\varepsilon_B-m_A-\omega=0,                                 
\end{equation}
where $\varepsilon_B=\sqrt{m^2_B+\p^2_B}$, $~\p_B=\q-\p_x$, $~\p_x=
\sqrt{\varepsilon^2_x-m^2}$, and $m_A$, $m_B$, and $m$ are masses of the 
target, recoil nucleus and nucleon, respectively. 
The missing momentum $p_m$ and missing energy $\varepsilon_m$ are defined by 
\begin{subequations}
\begin{align}
\label{Eq.5}
\p_m & = \p_x-\q
\\
\label{eps_m}
\varepsilon_m & = m + m_B - m_A                                           
\end{align}
\end{subequations}
From Eq.\eqref{Eq.4} the total energy of the ejected nucleon is given by
\begin{equation}\label{Eq.6}
\varepsilon_x=\omega + m_A - \varepsilon_B \approx \omega + m - 
(\varepsilon_m + p^2_m/2m_B)                                 
\end{equation}
and neglecting the recoil nucleus energy $p^2_m/2m_B$, the nucleon kinetic
energy can be written as 
\begin{equation}\label{Eq.7}
T_N=\omega - (\varepsilon_m + p^2_m/2m_B) \approx \omega - \varepsilon_m.     
\end{equation}
If we assume the target nucleon to be at rest inside a nucleus we have
\begin{equation}\label{Eq.8}
k_i=k_f\cos\theta_f + p_x\cos\theta_p,     
\end{equation}
where $k_f=\vert \k_f\vert$, $p_x=\vert \p_x \vert$, and $\cos\theta_f$ and
$\cos\theta_p$ are lepton and nucleon scattering angles, respectively. From
Eq.\eqref{Eq.8} it follows that the lepton and nucleon scattering angles are
connected by the relation
\begin{equation}\label{Eq.9}
  \cos\theta_f = \frac{k_i - p_x\cos\theta_p}{\var_i - T_N - \var_m}. 
\end{equation}

The leptonic tensor $L^{(nc)}_{\mu \nu}$ is separated into symmetric and 
antisymmetric components that are given as in Ref.~\cite{BAV2}. Note that 
the weak lepton NC is conserved for massless neutrino and $q^{\mu}
L^{(nc)}_{\mu\nu}=L^{(nc)}_{\mu\nu}q^{\nu}=0$. All the nuclear structure information 
and FSI effects are contained in the weak NC nuclear 
tensor $W^{(nc)}_{\mu \nu}$, which is given by the bilinear product of the 
transition matrix elements of the nuclear NC operator $J^{(nc)}_{\mu}$ between 
the initial nucleus state $|A\rangle$ and the final state $|B_f\rangle$ as 
\begin{eqnarray}
\label{Eq.10}
W^{(nc)}_{\mu \nu } &=& \sum_f \langle B_f,p_x\vert                           
J^{(nc)}_{\mu}\vert A\rangle \langle A\vert
J^{(nc) \dagger}_{\nu}\vert B_f,p_x\rangle,              
\label{W}
\end{eqnarray}
where the sum is taken over undetected states. 

General expressions for the cross sections of the exclusive and inclusive 
CCQE neutrino scattering off nucleus are given in Ref.~\cite{BAV2} in terms of 
weak response functions. For calculation of the NCE scattering this expression
 can be rewritten in the form
\begin{widetext}
\begin{align}\label{Eq.11}
\frac{d^5\sigma^{(nc)}}{d\var_f d\Omega_f d\Omega_x} &=
\frac{\vert\p_x\vert\varepsilon_x}{(2\pi)^5}G^2                         
\var_f \vert \k_f \vert R \big \{ v_0R^{(nc)}_0 + v_TR^{(nc)}_T
 + v_{TT}R^{(nc)}_{TT}\cos 2\phi + v_{zz}R^{(nc)}_{zz}
\notag \\
& +(v_{xz}R^{(nc)}_{xz} - v_{0x}R^{(nc)}_{0x})\cos\phi  
-v_{0z}R^{(nc)}_{0z} + h\big[v_{yz}(R^{\prime (nc)}_{yz}\sin\phi +
  R^{(nc)}_{yz}\cos\phi)
\notag \\
& - v_{0y}(R^{\prime (nc)}_{0y}\sin\phi + R^{(nc)}_{0y}\cos\phi) -
v_{xy}R^{(nc)}_{xy}\big]\big\},
\end{align}
\end{widetext}
where the response functions $R_i$ are suitable combinations of the hadron
tensor components $W^{(nc)}_{\mu\nu}$, and the coefficients $v_i$ are calculated
for massless neutrino. The exclusive cross section as a function of $\var_f$
and $\cos\theta_f$ can be obtained by integrating the exclusive
cross sections Eq.\eqref{Eq.11} over the azimuthal angle $\phi$ and missing
momentum $p_m$
\begin{eqnarray}\label{Eq.12}
\bigg(\frac{d^3\sigma^{(nc)}}{d\varepsilon_f d\Omega_f}\bigg)_{ex} &=&
\int_{0}^{2\pi}d\phi\int_{p_{min}}^{p_{max}}                                
dp_m\frac{p_m}{p_x \vert\q \vert}R_c
\times\frac{d^5\sigma^{(nc)}}{d\varepsilon_f d\Omega_f d\Omega_x},
\end{eqnarray}
where $p_m=\vert\p_m\vert,~ \p_m=\p_x-\q$, and
\begin{subequations}
\begin{align}
\cos\theta_x &= \frac{\p^2_x + \q^2 - \p^2_m}{2p_x\vert\q\vert},    
\\
R_c &= 1 + \frac{\varepsilon_x}{2p^2_x\varepsilon_B}
(\p^2_x +\q^2 - \p^2_m).
\end{align}
\end{subequations}
The integration limits $p_{min}$ and $p_{max}$ are given in Ref.~\cite{Petti}.
As the outgoing neutrino is undetected the differential cross section
Eq.\eqref{Eq.12} in ``the target nucleon at rest'' approximation
can be rewritten, using Eqs.\eqref{Eq.7}, and \eqref{Eq.8},
as function of $p_x$ and $\cos\theta_p$ as
\begin{equation}
  \label{Eq.13}
  \bigg(\frac{d^2\sigma^{(nc)}}{dp_x d\cos\theta_p}\bigg)_{ex} \approx R_p
  \bigg(\frac{d^2\sigma^{(nc)}}{d\var_f d\cos_f}\bigg)_{ex},
\end{equation}
where $R_p=p^2_x/[\var_x(\var_i-\omega)]$.


\subsection{Model}

We describe the neutrino-nucleon NCE scattering within the RDWIA approach.
This formalism is based on the impulse approximation (IA),
assuming that the incoming neutrino interacts with only one nucleon (which is 
subsequently emitted) while the remaining ($A$-1) nucleons in the target are 
spectators. The nuclear current is written as the sum of single-nucleon 
currents. Then the nuclear matrix element in Eq.\eqref{Eq.10} takes the form
\begin{eqnarray}\label{Eq.14}
\langle p,B\vert J^{\mu (nc)}\vert A\rangle &=& \int d^3r~ \exp(i\t\cdot\r)
\overline{\Psi}^{(-)}(\p,\r)
\Gamma^{\mu (nc)}\Phi(\r),                                          
\end{eqnarray}
where $\Gamma^{\mu (nc)}$ is the NC vertex function, $\t=\varepsilon_B\q/W$ is the
recoil-corrected momentum transfer, $W=\sqrt{(m_A+\omega)^2-\q^2}$ is the
invariant mass, $\Phi$ and $\Psi^{(-)}$ are the relativistic bound-state and
outgoing wave functions.

The single-nucleon neutral current has a $V{-}A$ structure $J^{(nc)\mu} = 
J^{\mu (nc)}_V + J^{\mu (nc)}_A$. For a free-nucleon vertex function, 
$\Gamma^{\mu (nc)} = \Gamma^{\mu (nc)}_V + \Gamma^{\mu (nc)}_A$, we use the vector 
current vertex function 
\begin{equation}\label{Eq.15}
\Gamma^{\mu (nc)}_V = F^{(nc)}_V(Q^2)\gamma^{\mu} + 
{i}\sigma^{\mu \nu}q_{\nu}F^{(nc)}_M(Q^2)/2m,                      
\end{equation}
and the axial current vertex function
\begin{eqnarray}\label{Eq.16}
\Gamma^{\mu (nc)}_A = F^{(nc)}_A(Q^2)\gamma^{\mu}\gamma_5 +         
F^{(nc)}_P(Q^2)q^{\mu}\gamma_5.
\end{eqnarray}

The vector form factors $F^{(nc)}_i$ ($i=V,M$) are related to the corresponding 
electromagnetic ones for protons $F^p_i$ and neutrons $F^n_i$, plus a possible 
isoscalar strange-quark contribution $F^s_i$~\cite{Alberico1}, i.e. 
\begin{subequations}
\begin{align}
\label{Eq.17}
F^{(nc)}_V & = \tau_3(0.5-\sin^2\theta_W)(F^p_1 - F^n_1)           
- \sin^2\theta_W(F^p_1 + F^n_1) - F^s_V/2
\\
F^{(nc)}_M & = \tau_3(0.5-\sin^2\theta_W)(F^p_2 - F^n_2) 
 - \sin^2\theta_W(F^p_2 + F^n_2) - F^s_M/2,                       
\end{align}
\end{subequations}
where $\tau_3=+(-1)$ for proton (neutron) knockout and $\theta_W$ is the 
Weinberg angle ($\sin^2\theta_W\approx 0.2313$). The axial $F_A^{(nc)}$ form 
factor is expressed as 
\begin{eqnarray}\label{Eq.18}
\Gamma^{\mu (nc)}_A = (\tau_3 F_A - F^s_A)/2,                      
\end{eqnarray}
where $F^s_A$ describes possible strange-quark contributions. This form factor
in the dipole approximation is parameterized as 
\begin{eqnarray}\label{Eq.19}
F^{(nc)}_A = \frac{1}{2}\frac{\tau_3F_A(0)-\Delta s}{(1+Q^2/M^2_A)^2},  
\end{eqnarray}
with $F_A(0)=1.272$, and $\Delta s$ describes the possible strange-quark
contribution. The contribution of the pseudoscalar form factor $F^{(nc)}_p$,
in Eq.~\eqref{Eq.16} is proportional to the mass of the scattered lepton,
so vanishes for neutral currents. 

Measurements of the strange vector form factors in parity violating electron
scattering point to small strangeness of $F^s_i$ form factors~\cite{Liu}.
Therefore in this work we 
neglect the strangeness contributions, i.e., it is supposed that 
$F^s_V=F^s_M=0$. For the nucleon form factors $F^{p(n)}_i$ the 
approximation of Ref.~\cite{MMD} is used. Because the bound nucleons are 
off-shell we employ the de~Forest prescription~\cite{deFor} and Coulomb gauge 
for the off-shell vector current vertex $\Gamma^{\mu}_V$.

In the RDWIA calculations the independent particle shell model (IPSM) is
assumed in the calculations of the nuclear structure.
In Eq.(\ref{Eq.14}) the relativistic bound-state wave functions for nucleons
$\Phi$ are obtained as the self-consistent solutions of relativistic Hartree
equations, derived within a relativistic mean-field approach
~\cite{Horowitz:1981xw, Horowitz:1991} with the normalization factors
$S_{\alpha}$ relative to the full occupancy of the
IPSM orbital $\alpha$ of ${}^{40}$Ca. For ${}^{40}$Ca and ${}^{40}$Ar an average
factor $\langle S \rangle \approx 87\%$. This estimation of depletion of hole
states follows from the RDWIA analysis of ${}^{40}$Ca$(e,e'p)$
data~\cite{BAV4}. The source of the reduction of the
$(e,e'p)$ spectroscopic factors with respect to the mean field values are 
the short-range and tensor correlations in the ground state, leading to the
appearance of the high-momentum and high-energy component in the nucleon
distribution in the target. Mean values of proton and neutron binding energies
and occupancies of shells are given also in Ref.~\cite{BAV4}.

In the RDWIA model, final state interaction effects for the outgoing nucleons
are taken into account.  The system of two coupled first-order Dirac equations
is reduced to a single second-order Schr\"odinger-like equation for the upper
component of the Dirac wave function $\Psi$. This equation contains a
phenomenological relativistic optical potential.
The optical potential consists of a real part, which describes the rescattering
of the ejected nucleon and an imaginary part which accounts for its absorption
into unobserved channels. The LEA program~\cite{LEA} is used
for the numerical calculation of the distorted wave functions with the EDAD1
parametrization~\cite{Cooper} of the relativistic optical potential for
calcium.

The RDWIA model was successfully tested in Ref.~\cite{BAV4} against $A(e,e'p)$
data for electron scattering off ${}^{40}$Ca. In Ref.~\cite{BAV5} it was shown that
this approach describes well the electron scattering data for carbon,
calcium, and argon at different kinematics. The calculated and measured
inclusive cross sections are in agreement within the experimental uncertainties.
 The RDWIA calculations are generally expected to be more accurate at higher
$Q^2$, since QE $(e,e'p)$ is expected to be dominanted by single-particle
interactions in this regime of four-momentum transfer, and two-body currents
stemming from meson-exchange currents are not needed to explain the data at this
$Q^2$~\cite{Fissum}.

\begin{figure*}
  \begin{center}
    \includegraphics[height=9cm,width=11cm]{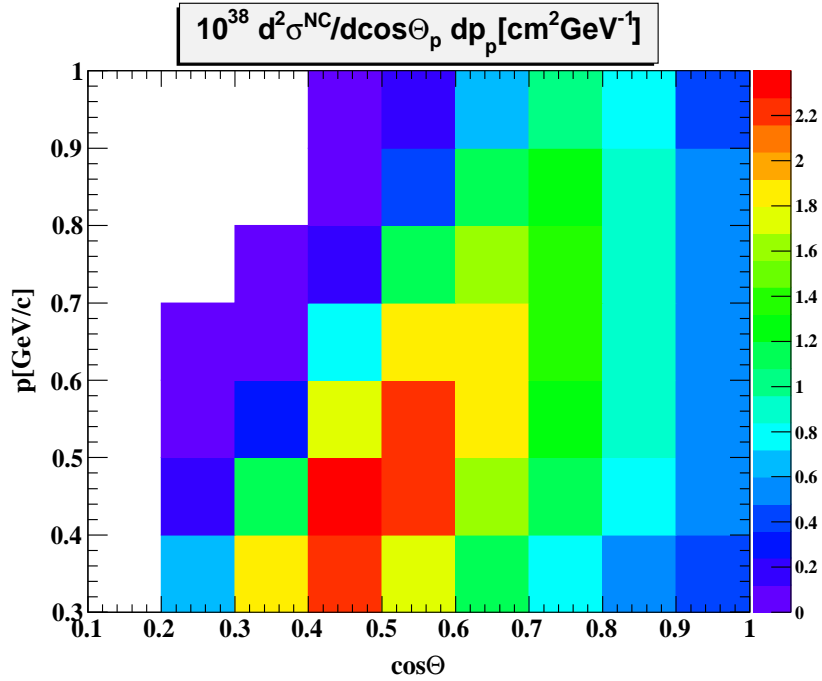}
  \end{center}
  \caption{\label{Fig1} The flux-integrated double-differential NCE
    semiexclusive cross section as a function of proton momentum and the cosine
    of the proton scattering angle.
  }
\end{figure*}

\section{Results and discussion}

\subsection{CCQE and NCE semiexclusive differential cross section}

The first measurement of exclusive CCQE-like flux-integrated cross
sections was performed using the MicroBooNE LArTPC neutrino detector 
presented in Ref.~\cite{MicroB3}. A specific subset of CCQE-like interactions
(CC1p0$\pi$ interactions) includes CC $\nu_{\mu}-{}^{40}Ar$ scattering events
with a detected muon and exactly one proton, with momenta greater than 100 and
300 MeV/c, respectively. The data were taken in a phase-space region that
corresponds to $0.1 < p_{\mu} < 1.5$ GeV/c, $0.3 < p_p < 1$ GeV/c,
$-0.65 < \cos\theta < 0.95$, and $\cos\theta_p > 0.15$. The MicroBooNE
detector is located along the Booster Neutrino Beam at Fermilab. The BNB
energy spectrum extends to 2 GeV and peaks around 0.7 GeV~\cite{BNB}.

For these CC1p0$\pi$ events the flux-integrated
$\nu_{\mu}-{}^{40}Ar$ double differential cross section in muon and proton
momenta and angles were measured, as a function of the calorimentric measured neutrino
energy and reconstructed momentum transfer. The flux-integrated cross section
is defined as
\begin{eqnarray}
\label{Eq.20}
\left\langle \frac{d\sigma}{dpd\cos\theta}(p,\cos\theta)\right\rangle &=&  
\int W_{\nu}(\varepsilon_i)
\frac{d\sigma}{dpd\cos\theta}(\varepsilon_i,p,\cos\theta) 
d\varepsilon_i,
\end{eqnarray}
where $W_{\nu}$ is a unit-normalized neutrino flux
\begin{equation}\label{Eq.21}
W_{\nu}(\var_i)=I_{\nu}(\var_i)/\Phi_{BNB}     
\end{equation}
and
\begin{equation}\label{Eq.22}
\Phi_{BNB}=\int I_{\nu}(\var_i)d \var_i     
\end{equation}
is determined by integration of the neutrino flux over $0 < \var_i < 3$ GeV. As
follows from~\eqref{Eq.20} the differential flux-integrated cross sections
depend on the shape of the neutrino spectrum.
\begin{figure*}
  \begin{center}
    \includegraphics[height=9cm,width=11cm]{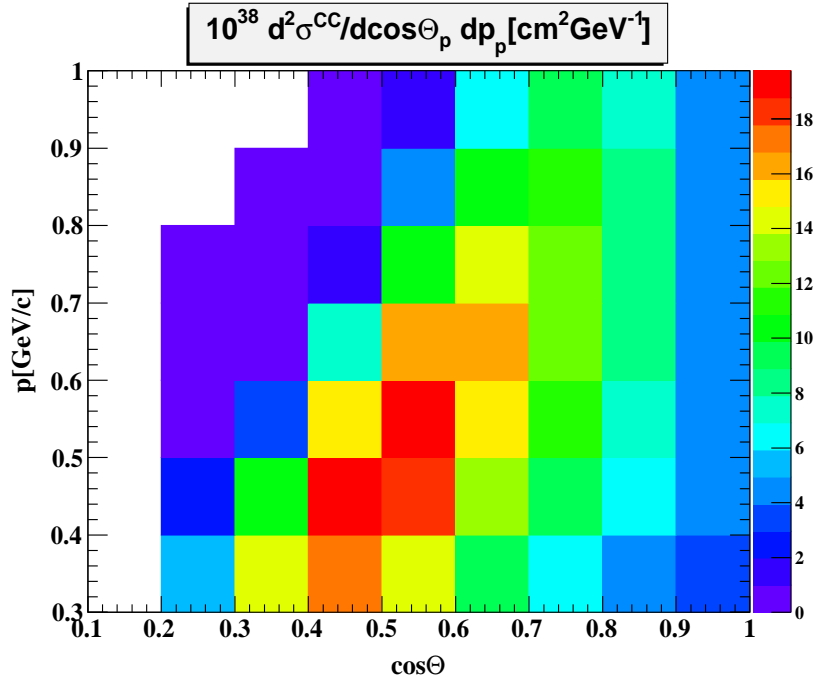}
  \end{center}
  \caption{\label{Fig2} Same as Fig.1, but for the CCQE semiexclusive
    reaction.
}
\end{figure*}
In Ref.~\cite{BAV6} the
flux-integrated CCQE semiexclusive cross sections for $\nu_{\mu}{}^{40}$Ar
scattering were calculated within the RDWIA model and compared with the
MicroBooNE data.

In this work we calculate within this approach the semiexclusive
neutral-current elastic scattering muon neutrino off argon.
A specific subset of this interaction includes signal events with a
detected one proton and no other particles (NC1p) in the final state. We don't
consider meson-exchange current and nucleon-nucleon pair contributions. The
calculations are performed with $M_A=1$ GeV and 1.2 GeV, taking into account
the MicroBooNE momentum threshold for protons, i.e. $0.3 \leq p_p \leq 1$ GeV/c
and $\cos\theta_p > 0.15$. The values $1 \leq M_A \leq 1.2 $ GeV are in
agreement with the best fit values $M_A=1.15 \pm 0.03$ GeV and
$M_A=1.2 \pm 0.06$ GeV obtained from the CCQE-like fit of the MiniBooNE and
MINERvA data in Refs.~\cite{Wilkinson:2016wmz, Butkevich:2018hll}. For modeling
electron and muon neutrinos in Ref.~\cite{MicroA} the ``MicroBooNE Tune'' value
of $M_A=1.1\pm0.1$ GeV is used in the GENIE generator, whereas the
post-ND280-fit value of $M_A=1.13\pm 0.08$ GeV is applied in the NEUT
model~\cite{T2KA}.

The flux-integrated double-differential cross sections
$d^2\sigma/dp_pd\cos\theta_p$ of the semiexclusive NCE and CCQE
$\nu_{\mu}-{}^{40}$Ar scattering are presented in Figs.1 and 2,
respectively as functions of proton momentum and scattering angle.
\begin{figure*}
  \begin{center}
    \includegraphics[height=16cm,width=16cm]{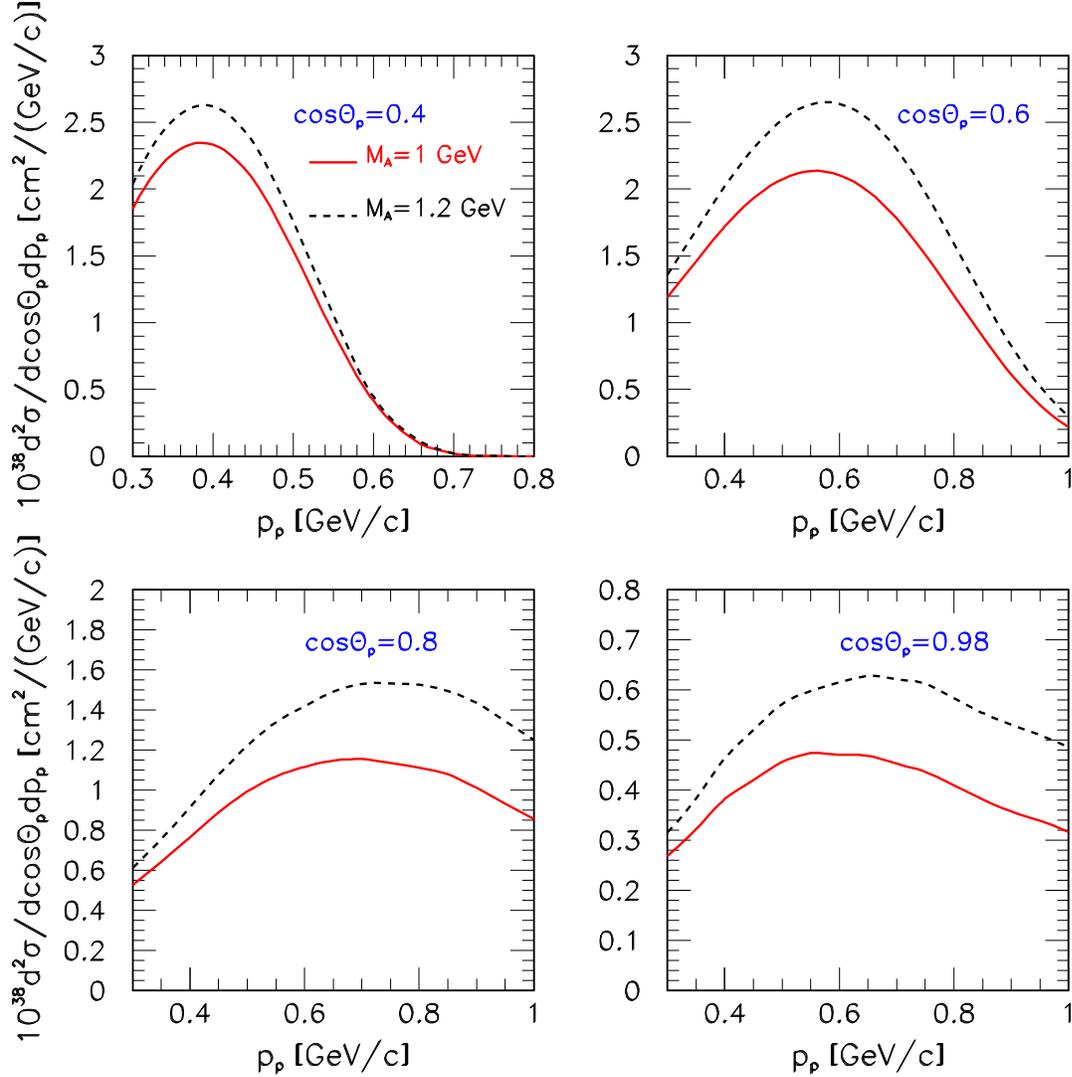}
  \end{center}
  \caption{\label{Fig.3} The flux-integrated semiexclusive NCE 
    $d\sigma^{nc}dp_p$ cross section for $\nu_{\mu}-{}^{40}Ar$
    scattering as a function of $p_{\mu}$ for the four proton scattering
    angles: $\cos\theta_p=0.4, 0.6, 0.8$ and 0.98. As shown in the key, cross
      sections were calculated with $M_A=1$ GeV and 1.2 GeV. 
} 
\end{figure*}
Here, the
results were obtained with the value of $M_A=1$ GeV
and $\Delta s=0$. The maximum of the calculated cross sections is in the
region $0.3 \leq p_p \leq 0.6$ GeV/c and $0.4 \leq \cos\theta_p \leq 0.6$.
Moreover, the shapes of the $d^2\sigma^{nc}/dp_pd\cos_p$ and
$d^2\sigma^{cc}/dp_pd\cos_p$ distributions are very similar.

Fig.3 shows the flux-integrated NCE $d\sigma^{nc}/dp_p$ cross
sections calculated with $\Delta s=0$ as a function of $p_p$ for several values
of the proton scattering angle and Fig.4 shows the
$d\sigma^{nc}/d\cos\theta_p$ cross section as a function of $\cos\theta_p$ for
several values of the proton momentum. On can observe from Fig.3 that in the
region of the of the NCE peak the cross sections calculated with $M_A=1.2$ GeV
at $\cos\theta_p=0.4$(0.98) are larger than ones calculated with $M_A=1$ GeV by
about 15\%(45\%).
\begin{figure*}
  \begin{center}
    \includegraphics[height=16cm,width=16cm]{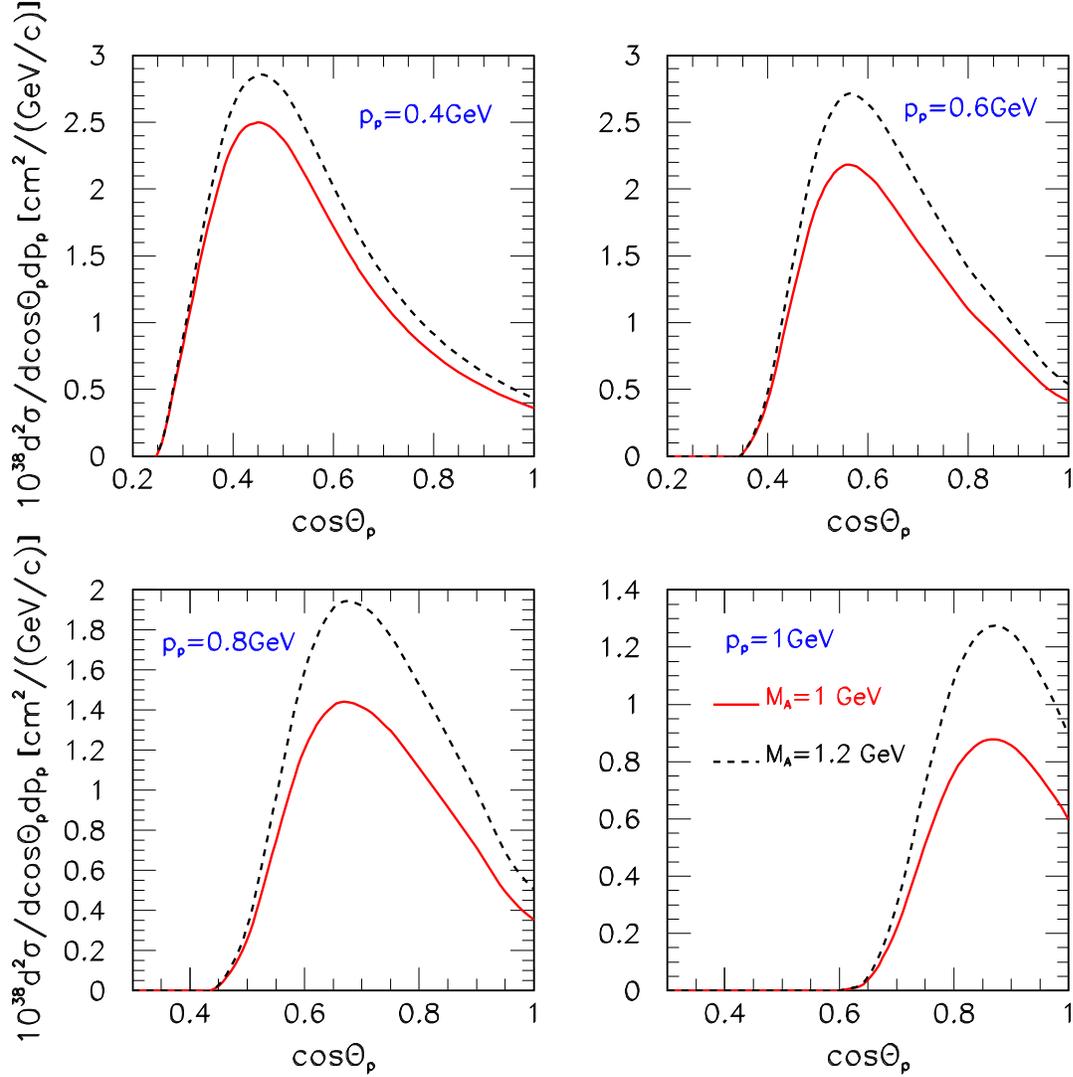}
  \end{center}
  \caption{\label{Fig.4} Same as Fig.3 but as a function of $\cos\theta_p$ for
    the four proton momenta: $p_p=0.4, 0.6, 0.8$, and 1 GeV/c.
} 
\end{figure*}
\begin{figure*}
  \begin{center}
    \includegraphics[height=15cm,width=15cm]{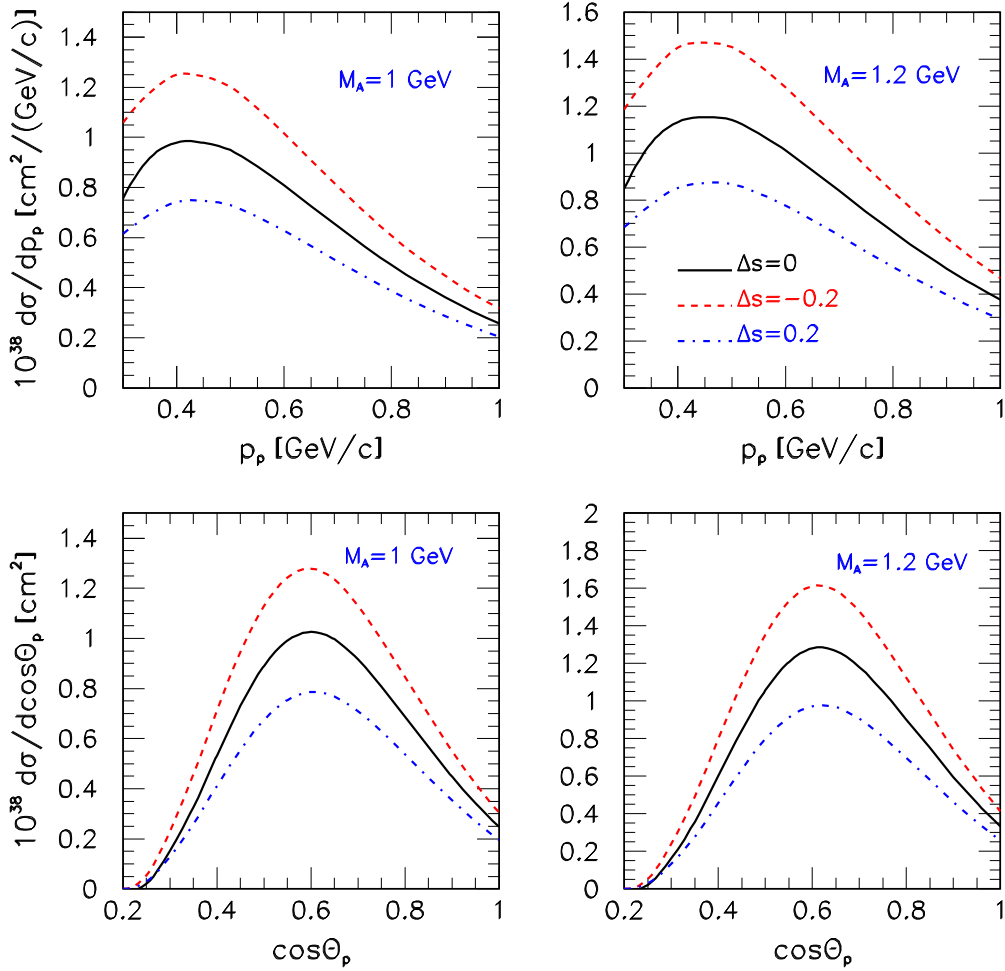}
  \end{center}
  \caption{\label{Fig.5} The flux-integrated differential $d\sigma^{nc}/dp_p$
    cross section (upper panel) as a function of proton momentum and
    $d\sigma^{nc}/d\cos\theta_p$  cross section (lower panel) as a function of
    $\cos\theta_p$. As shown in the key the semiexclusive cross sections were
    calculated with $M_A=1$ GeV and 1.2 GeV. Also shown is the strange quark
   effect on the NCE cross section with a value $\Delta s=-0.2$ (dashed line),
   $\Delta s=0$ (solid line), and $\Delta s=0.2$ (dashed-dotted line). 
}
\end{figure*}

The flux-integrated differential cross sections $d\sigma^{nc}/dp_p$ as a
function of proton momentum and $d\sigma^{nc}/d\cos\theta_p$ as a function of
the cosine of the proton scattering angle, calculated with $M_A=1$ GeV and
1.2 GeV are shown in Fig.5. The effect of a non-zero strange quark contribution
to the nucleon NC axial form factor also shown by comparing the results obtained
with $\Delta s=0$, $\Delta s=-0.2$, and $\Delta s =0.2$. These values
span the almost whole range of the values of $\Delta s$, extracted from
experimental data.
\begin{figure*}
  \begin{center}
    \includegraphics[height=15cm,width=12cm]{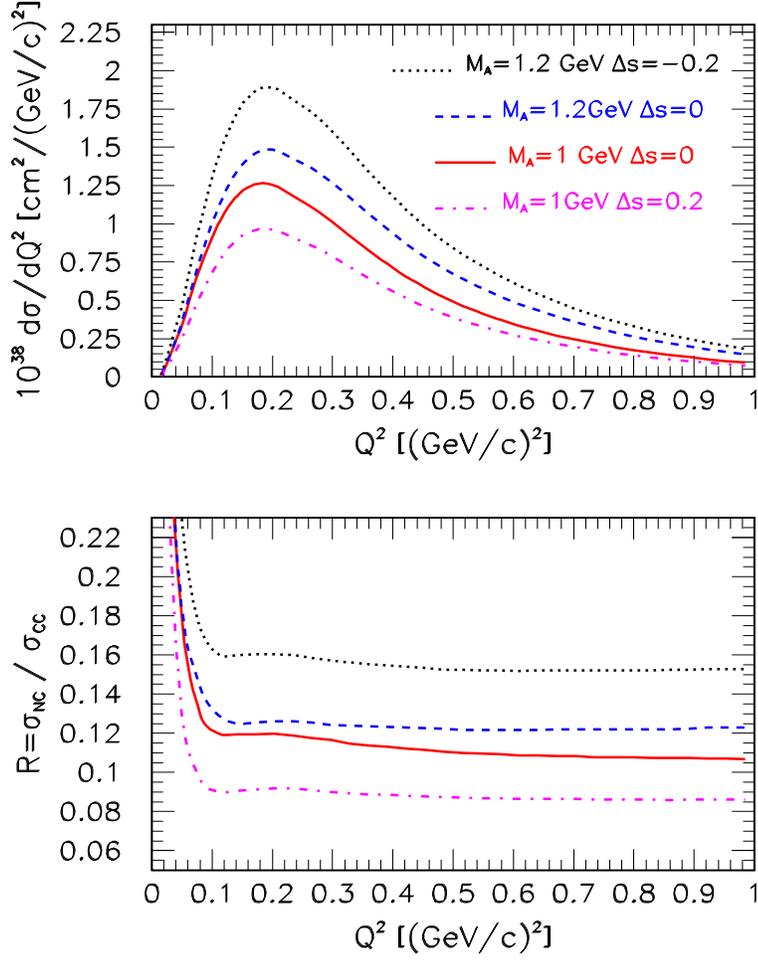}
  \end{center}
  \caption{\label{Fig.6} The flux-averaged $d\sigma^{nc}/dQ^2$ cross section
    (upper panel) for neutrino scattering on ${}^{40}$Ar and NCE/CCQE cross
    section ratio (lower panel) as a function of $Q^2$. The cross section and
    ratio are calculated with values of $M_A=1.2$ GeV and $\Delta s=-0.2$
    (dotted line); $M_A=1.2$ GeV and $\Delta s=0$ (dashed line); $M_A=1$ GeV
    and $\Delta s=0$ (solid line); $M_A=1$ GeV and $\Delta s =0.2$
    (dashed line). 
}
\end{figure*}
Note that the cross sections decrease when increasing
$\Delta s$. For example, at the NCE peak the cross sections are reduced by
about 65\% when $\Delta s$ running from $\Delta s=-0.2$ to $\Delta s=0.2$.

Figure 6 shows the flux-integrated differential cross section
$d\sigma^{nc}/dQ^2$ and a
$R=NCE/CCQE=\langle\sigma^{nc}/dQ^2\rangle/\langle d\sigma^{cc}/dQ^2\rangle$
cross section ratio as a function of $Q^2$, calculated with $M_A=1(1.2)$
GeV and $\Delta s=-0.2$, 0, and 0.2. The semiexclusive
$\langle d\sigma^{cc}/dQ^2\rangle$
cross section was calculated in Ref.~\cite{BAV6}. On can observe that the ratio
decreases slowly as $Q^2$ increases, and in the range of the NCE peak
the $d\sigma^{nc}/dQ^2$ cross section, calculated with $M_A=1.2$ GeV and
$\Delta s=-0.2$ is about two times larger then one obtained
with $M_A=1$ GeV and $\Delta s=0.2$. The NCE/CCQE ratio is used to search for
strangeness effects because the uncertainties of the absolute neutrino flux as
well as the sensitivity to the value of $M_A$ and nuclear effects are reduced
in this ratio. For example, in the range of the maximum the ratio
calculated with
$M_A=1.2$ GeV and $\Delta s=-0.2$ is about 1.7 times larger than the ratio
calculated with $M_A=1$ GeV and $\Delta s =0.2$. So, the theoretical
uncertainties on the $d\sigma^{nc}/dQ^2$ cross section and NCE/CCQE ratio due
to uncertainties of the values of $M_A$ and $\Delta s$ can reach 75-100\%.
A LArTPC detector's ability to detect low-energy
protons translates into ability to measure $F^s_A$ at low four-momentum
transfer where $F^s_A\sim \Delta s$. Neutrino-argon scattering experiments are a
suitable tool for extracting information about the contribution of strange
quark to the neutral current axial current.
\subsection{The flux-integrated cross sections and short base-line neutrino
 oscillations}
The CCQE and NCE signals are a two-body interaction with fully constrained
kinematics if the incoming and outgoing 4-vectors are known. In the LArTPC
detector the neutrino CCQE interaction products (one lepton and one proton)
can be accurately reconstructed. For these events, energy-momentum conservation
constraints allow the neutrino energy to be determined from the final-state
lepton energy and angle, the final-state proton energy and scattering angle,
or a combination of the final-state lepton and proton measurements. In
practice, the reconstructed kinematics of the CCQE and NCE events may suffer
substantial from smearing in the initial nucleon momentum, which is unknown, and
from final state interactions as the proton exits the nucleus.

We assume that the neutrino scatters off a single nucleon at rest and ignores
nuclear effects, including nucleon-nucleon correlations and nuclear recoil in
the quasielastic and elastic interactions. Then, the incoming neutrino energy
can be determined in the following ways: 
\begin{subequations}
\begin{align}
\label{Eq.23}
\var^l_{rec} & =\frac{\var_f(m-\epsilon_b)-(\epsilon_b^2+m^2_l-
  2m\epsilon_b)/2}{(m-\epsilon_b)-\varepsilon_f+k_f\cos\theta}      
\\
\var^p_{rec} & =\frac{T_p(m-\epsilon_b)-(m^2_l-\epsilon_b^2)/2}{p_x\cos\theta_p-
  (T_p+\epsilon_b)}                                                
\\
\var^{lp}_{rec} & = \var_f + T_p +\epsilon_b,                        
\end{align}
\end{subequations}
where $T_p$ is proton kinetic energy determined from the track length and
$\epsilon_b$ is the nucleon binding energy. For the NCE scattering, where the
outgoing neutrino is unmeasured, to reconstruct the incoming neutrino
energy we use the kinematics of the outgoing proton, i.e. Eq.(23b) with
$m_l=0$. In the case of well-reconstructed CCQE events $\var^l_{rec}$ and
$\var^p_{rec}$ will be in good agreement with $\var^{lp}_{rec}$.


We write the double differential cross section for the CCQE and NCE scattering
in terms of the final state proton momentum $p_p$ and reconstructed neutrino
energy $\var^p_{rec}$, using Eqs.~\eqref{Eq.13} and (23b)
\begin{eqnarray}\label{Eq.24}
  \frac{d^2\sigma}{dp_xd\var^p_{rec}} = R_{\var p} \frac{d^2\sigma}
       {dp_xd\cos\theta_p} ,                                             
\end{eqnarray}
where
\begin{eqnarray}\label{Eq.25}
  R_{\var p} = \frac{p^2_x+m^2_l-(T_x+\epsilon_b)^2}{2p_x(\var^p_{rec})^2}.     
\end{eqnarray}
Note that in Eq.~\eqref{Eq.25} $m_l=0$ for the case of NCE scattering. The
differential cross section as a function of $\var^p_{rec}$ is given by
\begin{eqnarray}\label{Eq.26}
  \frac{d\sigma}{d\var^p_{rec}} = \int_{p_{min}}^{p_{max}}\frac{d^2\sigma}
       {d\var^p_{rec}dp_x} dp_x,                                        
\end{eqnarray}
where $p_{min}=0.3$ GeV/c and $p_{max}=1$ GeV/c correspond a phase-space region
where the data were taken in the MicroBooNE experiment.

In the SBN program three detectors are used to measure the same neutrino
beam at different distances from the source. SBND, a 112 ton LArTPC is the
near detector will be located 110m downstream from the BND target and 60m
from the downstream face of the decay region to measure
the unoscillated neutrino flux. The far detector is the 476 ton active mass
ICARUS-T600 detector sited 600m from the target. The locations of the
near and far detectors are optimized for maximal sensitivity in search for
$\sim 1$ eV sterile neutrino~\cite{SBN}. In a sterile neutrino search based on
NCE interactions, the signal is the disappearance of any active neutrinos.
The NC disappearance provides the only means of directly constraining on the
admixture of mass state $\nu_4$ in the sterile flavor state $\nu_s$.

The SBN physics program also includes the study of neutrino-argon cross
sections. The multi-detector configuration allows simultaneous observation of
neutrino interaction at difference distances and independently measure at the
near and far detectors the flux-integrated CCQE and NCE cross sections as
functions of the reconstructed neutrino energy. Taking into account sterile
neutrino oscillation the flux-integrated cross section measured at the far
detector can be written as
\begin{eqnarray}\label{Eq.27}
\bigg(\frac{d\sigma^{(cc)(nc)}}{d\varepsilon^p_{rec}}\bigg)_{far} &=&
\int W_{\nu}(\var_i)P^{(cc)(nc)}_{(\nu_{\mu} \nu_{\mu})(\nu s)}(\var_i)          
\frac{d\sigma^{(cc)(nc)}}{d\var^p_{rec}}(\var_i,\var^p_{rec})d\var_i,
\end{eqnarray}
where $P^{(cc)}_{\nu_{\mu}\nu_{\mu}}$ and $P^{(nc)}_{\nu s}$ are probabilities of
survival of muon and active neutrino, respectively. If the unit-normalized
neutrino flux at the near detector $W_{\nu}^{ND}$ is the same as at the far
detector $W_{\nu}^{FD}$, the oscillation signal can be identified by observing
any variation in the ratio of the cross sections measured at the far and near
detectors, i.e.
\begin{eqnarray}\label{Eq.28}
  R = \bigg(d\sigma^{(cc)(nc)}/d\var^p_{rec}\bigg)_{far}\biggm/          
 \bigg(d\sigma^{(cc)(nc)}/d\var^p_{rec}\bigg)_{near}, 
\end{eqnarray}
where
\begin{eqnarray}\label{Eq.29}
\bigg(\frac{d\sigma^{(cc)(nc)}}{d\varepsilon^p_{rec}}\bigg)_{near} &=&
\int W_{\nu}(\var_i)                                                      
\frac{d\sigma^{(cc)(nc)}}{d\var^p_{rec}}(\var_i,\var^p_{rec})d\var^p_i
\end{eqnarray}
is the cross section measured at the near detector.

In the SBN experiment $W_{\nu}^{ND}$ at the near SBND is not exactly the same as
$W_{\nu}^{FD}$ at the ICARUS detector due to neutrino flux divergence~\cite{SBN}.
To search for the oscillation effects we need use the ratio of the measured and
predicted cross section at the far detector
\begin{eqnarray}\label{Eq.30}
  R^{exp}_{\sigma} = \bigg(d\sigma^{(cc)(nc)}/d\var^p_{rec}\bigg)^{data}_{FD}\biggm/  
 \bigg(d\sigma^{(cc)(nc)}/d\var^p_{rec}\bigg)^{pred}_{FD}.     
\end{eqnarray}
The predicted cross section can be expressed as
\begin{eqnarray}\label{Eq.31}
  \bigg(d\sigma^{(cc)(nc)}/d\var^p_{rec}\bigg)^{pred}_{FD}= F(\var^p_{rec})
  \bigg(d\sigma^{(cc)(nc)}/d\var^p_{rec}\bigg)^{cal}_{FD},           
\end{eqnarray}
where $\bigg(d\sigma^{(cc)(nc)}/d\var^p_{rec}\bigg)^{cal}_{FD}$ is the
flux-integrated cross section at the far detector calculated with
no-oscillation and $W_{\nu}=W^{FD}_{\nu}$. Discrepancy between
measured and calculated with $W_{\nu}=W^{ND}_{\nu}$ cross sections
at the near detector~\eqref{Eq.29}
\begin{eqnarray}\label{Eq.32}
  F(\var^p_{rec}) = \bigg(d\sigma^{(cc)(nc)}/d\var^p_{rec}\bigg)^{data}_{ND}\biggm/  
 \bigg(d\sigma^{(cc)(nc)}/d\var^p_{rec}\bigg)^{cal}_{ND}          
\end{eqnarray}
is extrapolated to produce the predicted cross section at the far detector for
the some oscillation hypothesis
\begin{eqnarray}\label{Eq.33}
  \bigg(d\sigma^{(cc)(nc)}/d\var^p_{rec}\bigg)^{pred}_{osc}= F(\var^p_{rec})
  \bigg(d\sigma^{(cc)(nc)}/d\var^p_{rec}\bigg)^{cal}_{osc},           
\end{eqnarray}
where $\bigg(d\sigma^{(cc)(nc)}/d\var^p_{rec}\bigg)^{cal}_{osc}$ is given by
Eq.\eqref{Eq.27} with $W_{\nu}=W^{FD}_{\nu}$. Then the ratio
\begin{eqnarray}\label{Eq.34}
  R_{\sigma} = \bigg(d\sigma^{(cc)(nc)}/d\var^p_{rec}\bigg)^{cal}_{osc}
  \biggm/ \bigg(d\sigma^{(cc)(nc)}/d\var^p_{rec}\bigg)^{cal}_{FD}     
\end{eqnarray}
can be used to determine significance of the muon and active neutrino
disappearance observed at the far detector, but it not suitable for the
determination of the oscillation parameters, since this ratio is not a function
 of the true neutrino energy.
\subsection{Oscillation model}
We use the 3+1 neutrino framework with an extended 4$\times$4 unitary
Pontekorvo-Maki-Nakagava-Sakata (PMNS) matrix $[U_{\alpha i}]$. The flavor
$\nu_{\alpha}$ and mass $\nu_i$ states are now connected by PMNS matrix
$\nu_{\alpha}=\sum U_{\alpha i}\nu_i$. Assuming the fourth neutrino mass
eigenstates is much heavier than the others $(m_4\gg m_3, m_2, m_1)$ the
short-baseline survival probability for muon neutrino takes the form 
\begin{eqnarray}\label{Eq.30}
  P_{\nu_{\mu}\rightarrow\nu_{\mu}} = 1 - \sin^22\theta_{\mu \mu}\sin^2\Delta_{41}   
\end{eqnarray}
and for active neutrinos
\begin{eqnarray}\label{Eq.31}
  P_{\nu s} = 1 - \sin^22\theta_{\mu s}\sin^2\Delta_{41},   
\end{eqnarray}
where $\Delta_{41}=\Delta m^2_{41}L/4E=1.267(\Delta m^2_{41}/eV^2)(GeV/E)(L/km)$
and $\theta_{\alpha \beta}$ is defined as the effective mixing angle.
These angles are expressed in terms of the matrix elements as
\begin{figure*}
  \begin{center}
    \includegraphics[height=13cm,width=18cm]{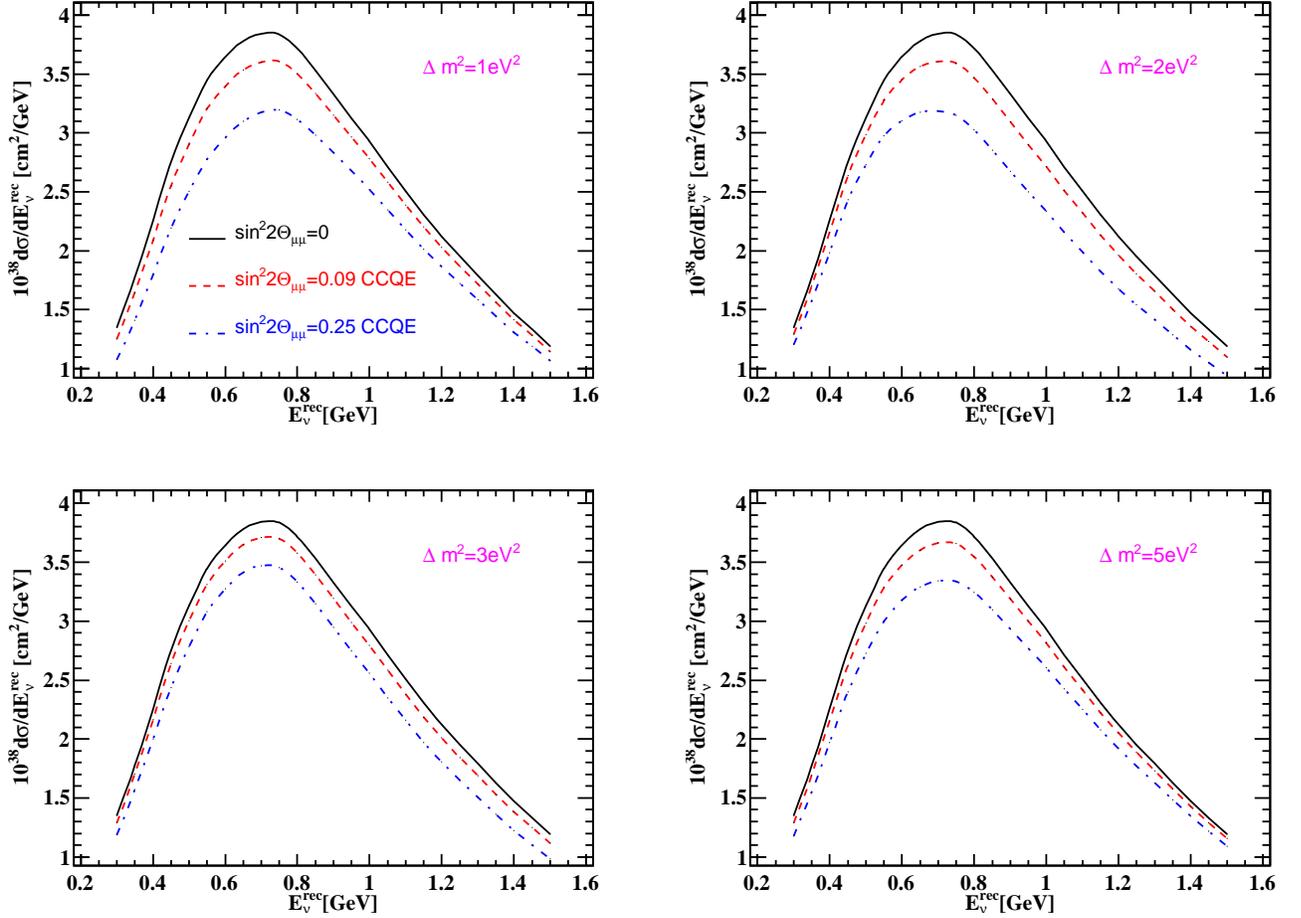}
  \end{center}
  \caption{\label{Fig.7} The flux-integrated semiexclusive CCQE cross section
    as a function of $E^{rec}_{\nu}$ for the four values of $\Delta m^2_{41}=1,2,3,$
    and 5 eV${^2}$. The solid line is result obtained with
    $\sin^22\theta_{\mu \mu}=0$, whereas the dashed and dash-dotted lines are
    results for $\sin^22\theta_{\mu \mu}=0.05$ and 0.25, respectively.
} 
\end{figure*}
\begin{subequations}
\begin{align}
\label{Eq.32}
\sin^22\theta_{\mu \mu} & = 4(1-\vert U_{\mu 4}\vert^2)\vert U_{\mu 4}\vert^2  
\\
\sin^22\theta_{\mu s} & = 4\vert U_{\mu 4}\vert^2\vert U_{s 4}\vert^2.       
\end{align}
\end{subequations}
The effective mixing angle $\sin^22\theta_{\mu s}$ can be related to other mixing
angles by imposing unitarity on the $4\times 4$ PMNS matrix
\begin{eqnarray}\label{Eq.33}
  \sum_{i=e,\mu,\tau,s} \vert U_{i4}\vert^2 = 1.                           
\end{eqnarray}
\begin{figure*}
  \begin{center}
    \includegraphics[height=13cm,width=18cm]{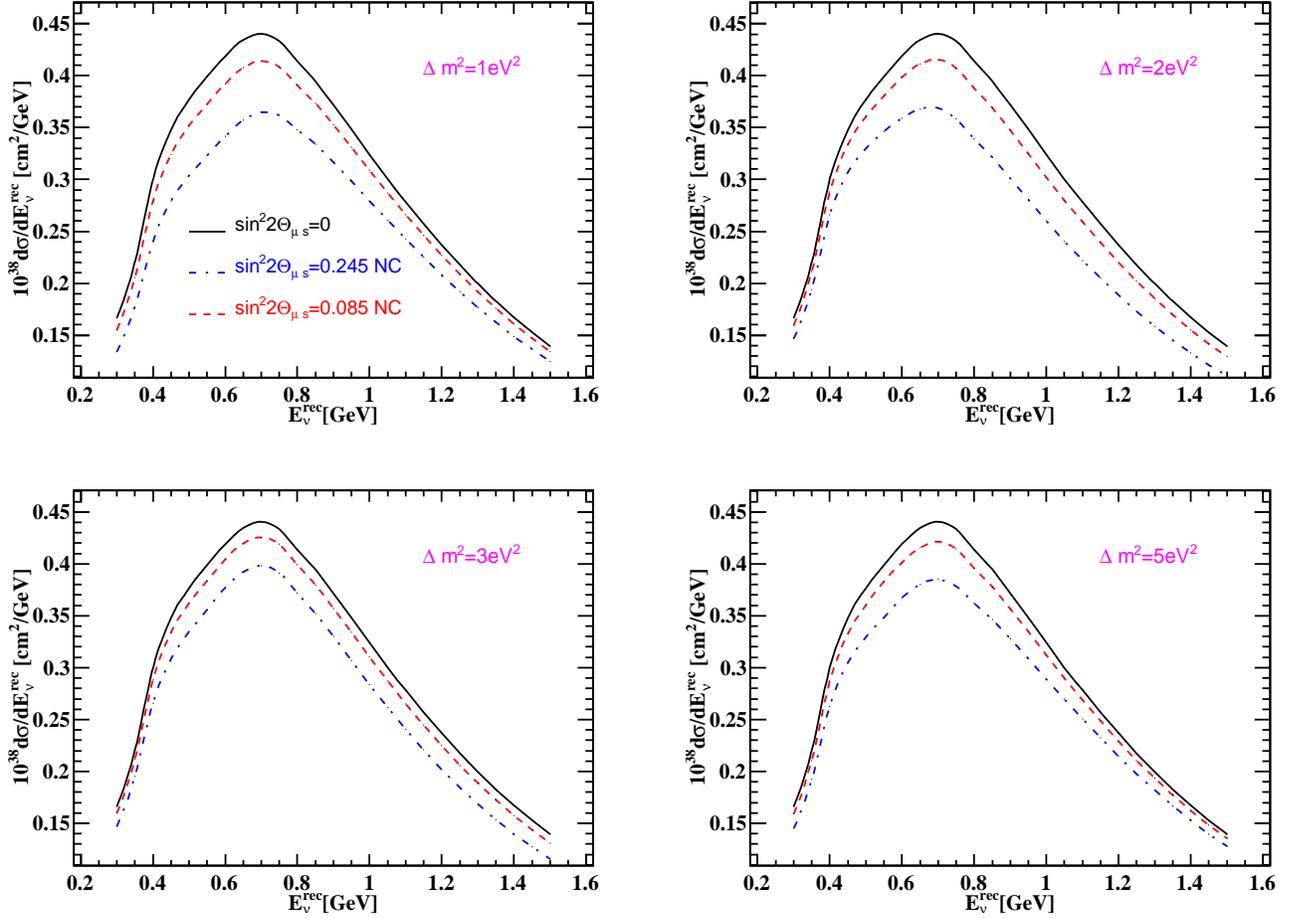}
  \end{center}
  \caption{\label{Fig.8} Same as Fig.7, but for NCE cross section and 
   $\sin^22\theta_{\mu s}=0$, 0.085, and 0.245.
} 
\end{figure*}
This equation relates the effective angles
\begin{eqnarray}\label{Eq.34}
  \sin^22\theta_{\mu s}=\sin^22\theta_{\mu \mu}-\sin^22\theta_{\mu e}-
  \sin^22\theta_{\mu \tau}                         
\end{eqnarray}
and provides a constraint $\sin^22\theta_{\mu \mu} \geq \sin^22\theta_{\mu s}$. 
To study sensitivities of the flux-integrated differential CCQE and NCE
cross sections to neutrino oscillations we use the allowed values of
$(\Delta m^2_{41}, \sin^22\theta_{\mu \mu})$ found in the MicroBooNE experiment
~\cite{Note106}, i.e. $1\leq\Delta m^2_{41}\leq 5$ eV$^2$ and
$0.09\leq\sin^22\theta_{\mu\mu}\leq 0.25$. In this experiment the 3+1 model was
tested with data using CCQE 1$\mu$1p events and muon neutrino disappearance was
not observed. On the other hand, to our knowledge, a global analysis of
sterile neutrino induced NC disappearance does not exist. Therefore, we
estimated the range of the allowed values of $\sin^22\theta_{\mu s}$ as
$0.085\leq\sin^22\theta_{\mu s}\leq 0.245$, using Eq.~\eqref{Eq.34} and assuming
that $\sin^22\theta_{\mu \tau}+\sin^22\theta_{\mu e} < 0.05$~\cite{Note116}.
\begin{figure*}
  \begin{center}
    \includegraphics[height=13cm,width=18cm]{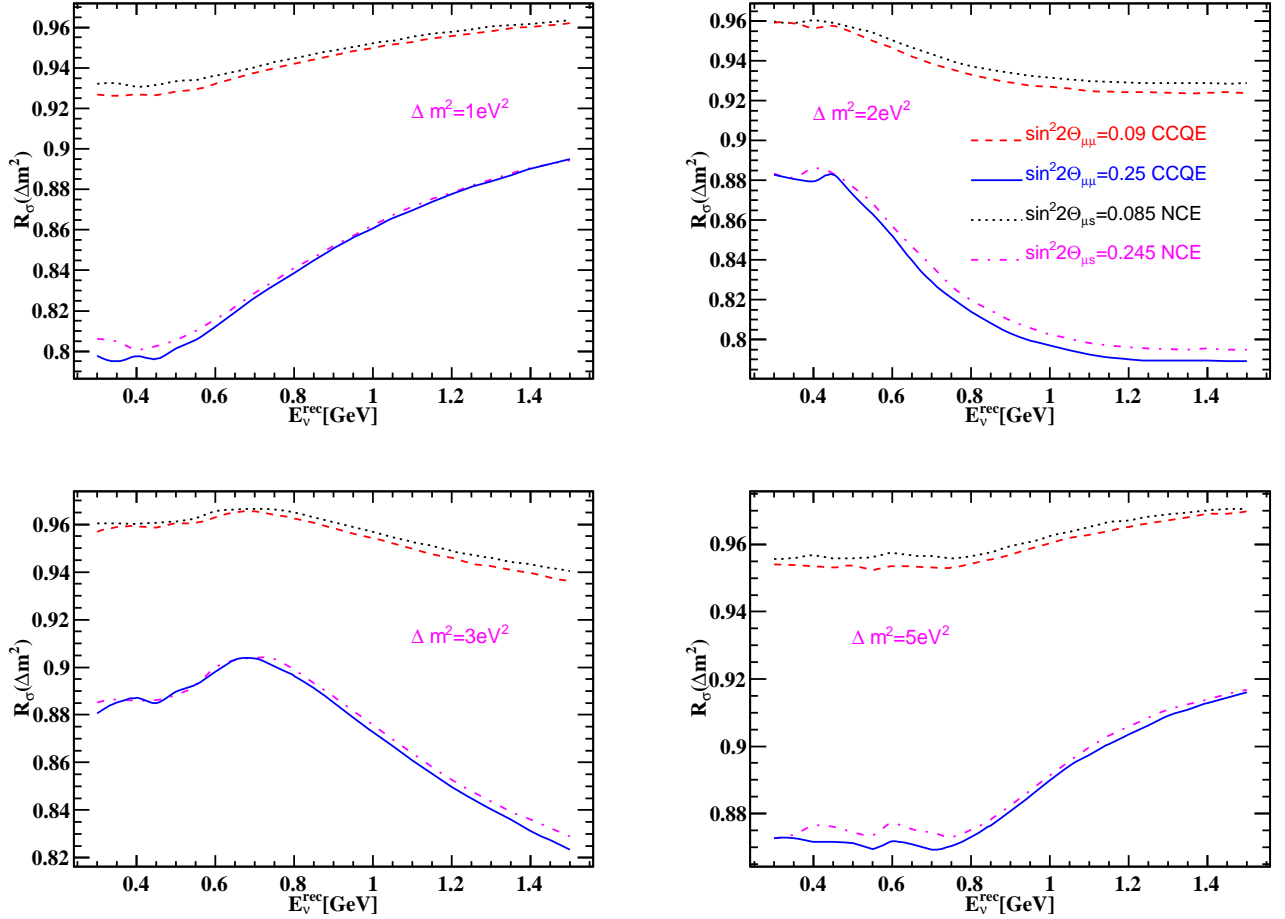}
  \end{center}
  \caption{\label{Fig.9} The ratio $R_{\sigma}$ of the flux-integrated
    semiexclusive CCQE and NCE cross sections as a function of $E^{rec}_{\nu}$.
    The cross sections at the far detector are calculated
    for the four values of $\Delta m^2_{41}=1,2,3,$ and 5 eV${}^2$. The solid and
    dashed lines are results obtained for CCQE interactions with
    $\sin^22\theta_{\mu \mu}=0.09$ and 0.25, respectively. The dotted and
    dash-dotted lines are results obtained for NCE interactions with
    $\sin^22\theta_{\mu s}=0.085$ and 0.245, respectively.
} 
\end{figure*}
\subsection{Sensitivity of the flux-averaged $d\sigma/d\var^p_{rec}$ cross
  sections  to the short base-line neutrino oscillations}

The flux-integrated semiexclusive differential CCQE and NCE cross sections of
$\nu_{\mu}{}^{40}$Ar scattering are presented in Figs.7 and 8, respectively. 
The figures show $d\sigma/dE^{rec}_{\nu}$ cross sections as functions of
reconstructed neutrino energy $E^{rec}_{\nu}=\var^p_{rec}$ (23c), calculated with
$M_A=1$ GeV and $\Delta s=0$. Here, the results obtained for the near detector
with null oscillation effects (i.e., with
\begin{figure*}
  \begin{center}
    \includegraphics[height=13cm,width=18cm]{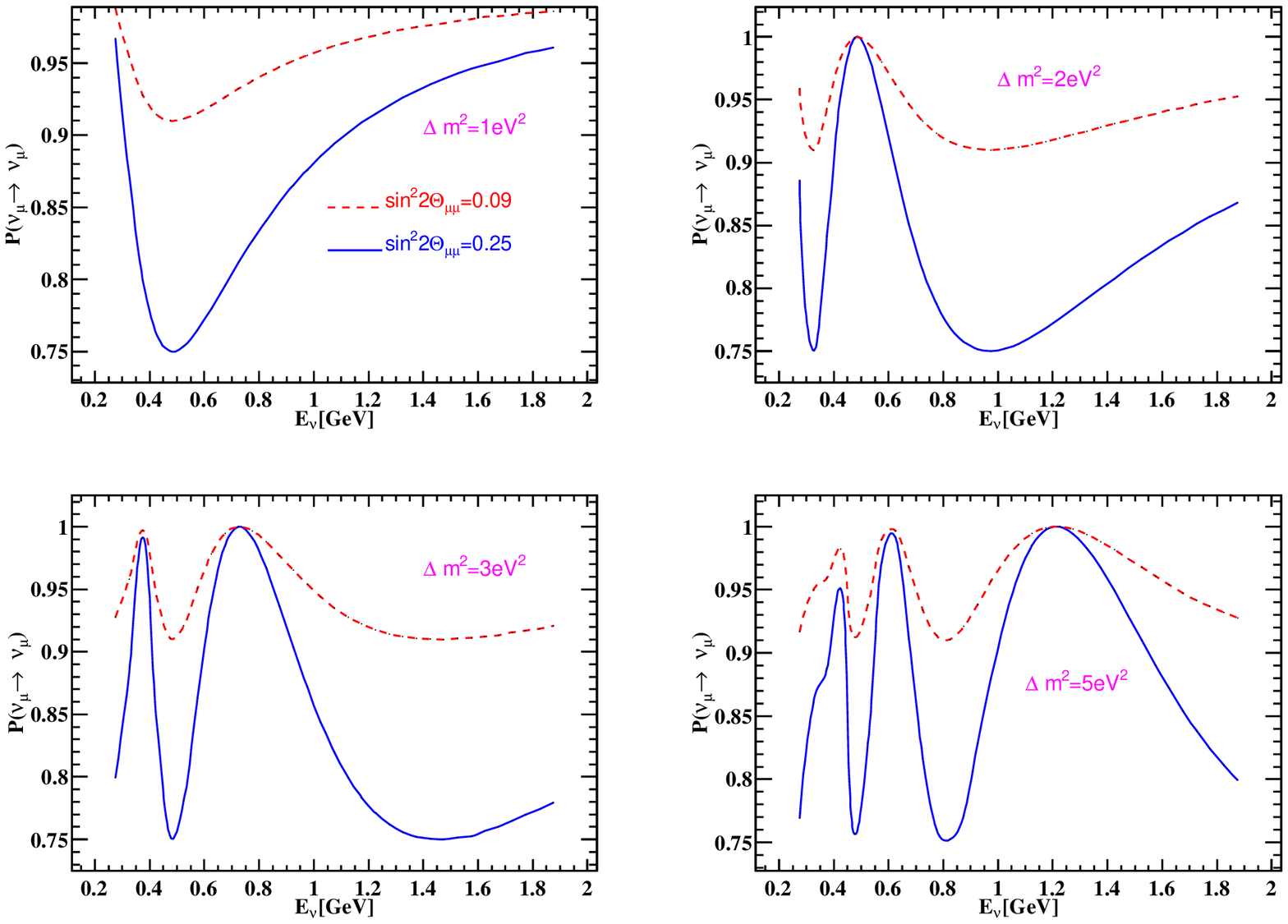}
  \end{center}
  \caption{\label{Fig.10} Survival probability for muon neutrino as a function
    of neutrino energy, calculated at $L=600$m for the values of
    $\Delta m^2_{41}=1,2,3,$ and 5 eV${}^2$. The solid and dashed lines are
    results
    obtained with $\sin^22\theta_{\mu \mu}=0.25$ and 0.09, respectively.
} 
\end{figure*}
$\sin^22\theta_{\mu \mu}=\sin^22\theta_{\mu s}=0$) are presented by the solid
lines. Also shown in these figures are
the cross sections, calculated for the far detector at the distance $L=600$m,
with the oscillation parameters $\Delta m^2_{41}=1,2,3$, and 5 eV${^2}$,
$\sin^22\theta_{\mu \mu}=0.09$, 0.25, $\sin^22\theta_{\mu s}=0.085$, 0.245.
Note that the maximum of the cross sections occurs at the same energy
$\var_i\simeq 0.7$ GeV as in BNB.
To show the effect of muon and active neutrino disappearance we
present in Fig.9 the ratios $R_{\sigma}$ of the cross sections~\eqref{Eq.34},
 as a function of $E^{rec}_{\nu}$.
This figure demonstrates that the form of the $R_{\sigma}(E^{rec}_{\nu})$
dependence is sensitive to the value of $\Delta m^2_{41}$. In Fig.10 the
survival probability for muon neutrino $P_{\nu_{\mu} \nu_{\mu}}$ at the distance
$L=600$ m, calculated for the same values of $\Delta m^2_{41}$ and
$\sin^22\theta_{\mu \mu}$ is shown as a function of neutrino energy $\var_i$,
for comparison. One can observe from these figures that the position of the
minimum
$\var^{rec}_{min}$ and maximum $\var^{rec}_{max}$ in the $R_{\sigma}(E^{rec}_{\nu})$
distribution depends on the value of  $\Delta m^2_{41}$ and correlates strongly
with the values of the energy, that correspond to the first minimum
$E_{min}=2.57\Delta m^2_{41}L/\pi$ and maximum $E_{max}=E_{min}/2$ in the
$P_{\nu_{\mu} \nu_{\mu}}$. For example, $\var^{rec}_{min}(E_{min})\approx 0.45(0.48)$
GeV at $\Delta m^2_{41}=1$ eV${}^2$; $\var^{rec}_{min}(E_{min})\approx 1.25(1.0)$
GeV, $\var^{rec}_{max}(E_{max})\approx 0.45(0.48)$ GeV at $\Delta m^2_{41}=2$
eV${}^2$; $\var^{rec}_{min}(E_{min})\approx 1.5(1.44)$ GeV,
$\var^{rec}_{max}(E_{max})\approx 0.75(0.75)$ GeV at $\Delta m^2_{41}=3$ eV${}^2$.
In the region of $\var^{rec}_{min}$ the effect of neutrino oscillation is
predicted to be $0.8\leq R_{\sigma}\leq 0.96$.
                 
Statistics of the CCQE and NCE-like candidate events collected at these
detectors will allow the dertimation of $R_{\sigma}$ ratios with high precision.
Therefore they can be used to search for the short base-line oscillations
for $\sim 1$ eV sterile neutrino, since with $\Delta m_{41}^2>2$ eV$^2$ it is
necessary to take into account the effects of oscillations at the near
detector.                    

\section{Conclusions}

In this article we study semiexclusive CCQE and NCE neutrino scattering on
argon in the framework of the RDWIA approach. We calculate the flux-integrated
differential cross sections with $M_A=1$ GeV and 1.2 GeV. The elastic scattering
cross sections are also evaluated with different strange quark contributions
to the NC axial form factor. It is shown that the maxima of the
$d^2\sigma^{cc}/dp_pd\cos\theta_p$ and $d^2\sigma^{nc}/dp_pd\cos\theta_p$ are in
the region $0.3\leq p_p\leq 0.6$ GeV/c and $0.4\leq\cos\theta_p\leq 0.6$.
Moreover, the shapes of these distributions are very similar. We calculate
the flux-integrated NCE $d\sigma^{nc}/ dp_p$, $d\sigma^{nc}/d\cos\theta_p$,
and $d\sigma^{nc}/dQ^2$ cross sections, as well as the NCE/CCQE ratio with
$\Delta s=-0.2, 0, 0.2$. Theoretical uncertainties in these cross sections
and ratio due to uncertainties in the NC axial form factor (in $M_A$ and
$\Delta s$) can reach 75-100\%.

The flux-integrated semiexclusive differential CCQE and NCE cross sections as 
functions of the reconstructed neutrino energy are calculated for the far
detector of the SBN experiment with no-oscillation and taking into account the 
short base-line sterile neutrino oscillation effects leading to the
disappearance of $\nu_{\mu}$ and $\nu_{active}$. We use the 3+1 framework  with
the values of oscillation parameters $1\leq \Delta m^2_{41} \leq 5$ eV${}^2$ and
$0.09(0.085)\leq\sin^22\theta_{\mu\mu (\mu s)}\leq 0.25(0.245)$. To show the
oscillation effects we have also calculated the $R_{\sigma}$ ratio of the cross
sections at the far detector. We found that the positions of
minimum and maximum of the $R_{\sigma}$ ratio depend on the value of
$\Delta m^2_{41}$ and correlate with the positions of the first minimum and
maximum in the survival probability for muon and active neutrino at the far
detector.

Therefore, the ratio of the measured and predicted cross sections at the far
detector of the SBN experiment can be used in a sterile-based oscillation
study.

\section*{Acknowledgments}

The author greatly acknowledges A. Habig for fruitful discussions and a
critical reading of the manuscript. I would like to thank A. Olshevskiy,
V. Naumov, O. Samoylov, N. Anfimov, I. Kokorin, and L. Kolupaeva for their
constructive comments and suggestions.

\end{document}